# Wide Field Hard X-ray Survey Telescope: *ProtoEXIST1*

J. Hong[a,*], J. E. Grindlay[a], N. Chammas[a], B. Allen[a], A. Copete[a], B. Said[a], M. Burke[a],
J. Howell[a], T. Gauron[a], R. G. Baker[b], S. D. Barthelmy[b], S. Sheikh[b], N. Gehrels[b],
W. R. Cook[c], J. A. Burnham[c], F. A. Harrison[c], J. Collins[d], S. Labov[d], A. Garson III[e],
and H. Krawczynski[e]

[a]Harvard Smithsonian Center for Astrophysics, 60 Garden St., Cambridge, MA 02138,
[b]NASA Goddard Space Flight Center, Greenbelt, MD 20771
[c]California Institute of Technology, Pasadena, CA 91125
[d]Lawrence Livermore National Laboratory, Livermore, CA 94550
[e]Washington University, St. Louis, MO, 63130

## ABSTRACT

We report our progress on the development of pixellated imaging CZT detector arrays for our first-generation balloon-borne wide-field hard X-ray (20 – 600 keV) telescope, *ProtoEXIST1*. Our *ProtoEXIST* program is a pathfinder for the High Energy Telescope (HET) on the Energetic X-ray Imaging Survey telescope (*EXIST*), a proposed implementation of the Black Hole Finder Probe. *ProtoEXIST1* consists of four independent coded-aperture telescopes with close-tiled (~0.4 mm gaps) CZT detectors that preserve their 2.5mm pixel pitch. Multiple shielding/field-of-view configurations are planned to identify optimal geometry for the HET in *EXIST*. The primary technical challenge in *ProtoEXIST* is the development of large area, close-tiled modules of imaging CZT detectors (1000 cm$^2$ for *ProtoEXIST1*), with all readout and control systems for the ASIC readout vertically stacked. We describe the overall telescope configuration of *ProtoEXIST1* and review the current development status of the CZT detectors, from individual detector crystal units (DCUs) to a full detector module (DM). We have built the first units of each component for the detector plane and have completed a few Rev2 DCUs (2x2 cm$^2$), which are under a series of tests. Bare DCUs (pre-crystal bonding) show high, uniform ASIC yield (~70%) and ~30% reduction in electronics noise compared to the Rev1 equivalent. A Rev1 DCU already achieved ~1.2% FWHM at 662 keV, and preliminary analysis of the initial radiation tests on a Rev2 DCU shows ~ 4 keV FWHM at 60 keV (vs. 4.7 keV for Rev1). We therefore expect about ≤1% FWHM at 662 keV with the Rev2 detectors.

## 1. INTRODUCTION

*ProtoEXIST1* is a first-generation balloon-borne wide-field hard X-ray (20 – 600 keV) telescope experiment, employing the coded-aperture imaging technique with CZT detectors. It is a technology pathfinder for the High Energy Telescope (HET) in the Energetic X-ray Imaging Survey telescope (*EXIST*) [1, 2]. *EXIST* is a proposed implementation of the Black Hole Finder Probe under NASA's Beyond Einstein Program. *EXIST* will detect and identify (possibly through follow-up observations with longer wavelength telescopes) highly obscured black holes, highly red-shifted Gamma-Ray Bursts (GRBs) occurring from births of possibly the very first black holes and other numerous interesting X-ray sources in the sky. To achieve these ambitious scientific goals, *EXIST* is envisioned to have two types of coded-aperture telescope arrays, which constitute a High Energy Telescope (HET: 10 – 600 keV) and a Low Energy Telescope (LET: 3 – 30 keV). The HET requires a very large array of detector area (~6 m$^2$) with relatively fine detector pixel size (~1.2 mm), good energy resolution (<3 keV) and high sensitivity over the ~10 – 600 keV band.

As a technology test bench for the HET in *EXIST*, the primary goals for our *ProtoEXIST* program are the following. First, demonstrate the technology of large area, close-tiled modules of imaging CZT detectors that can operate under the strict resource constraints (e.g. <100 μW/pixel power consumption). Second, determine the background properties on CZT detectors in near space environment and identify the optimal shielding configuration for the HET. Third, demonstrate high performance of *scanning* coded-aperture telescopes [3, 4]. The last item is very important for both the HET and LET to achieve the required wide dynamic range of sensitivity (~$10^4$–$10^5$) under unknown systematics (a few %), which is typically expected in large area detector systems as in *EXIST* [3, 4].

---
[*] Send correspondence to J. Hong (jaesub@head.cfa.harvard.edu). For easier reading, see the color version of this paper posted on http://arxiv.org/archive/astro-ph or at http://hea-www.harvard.edu/ProtoEXIST/.

Over the past few years, we have made great progress in development of *ProtoEXIST*. In the following, we describe the overall telescope configurations for *ProtoEXIST1*, which is mainly driven by the second and third objectives of *ProtoEXIST*. We also report the latest progress of the detector plane development.

## 2. TELESCOPE CONFIGURATION

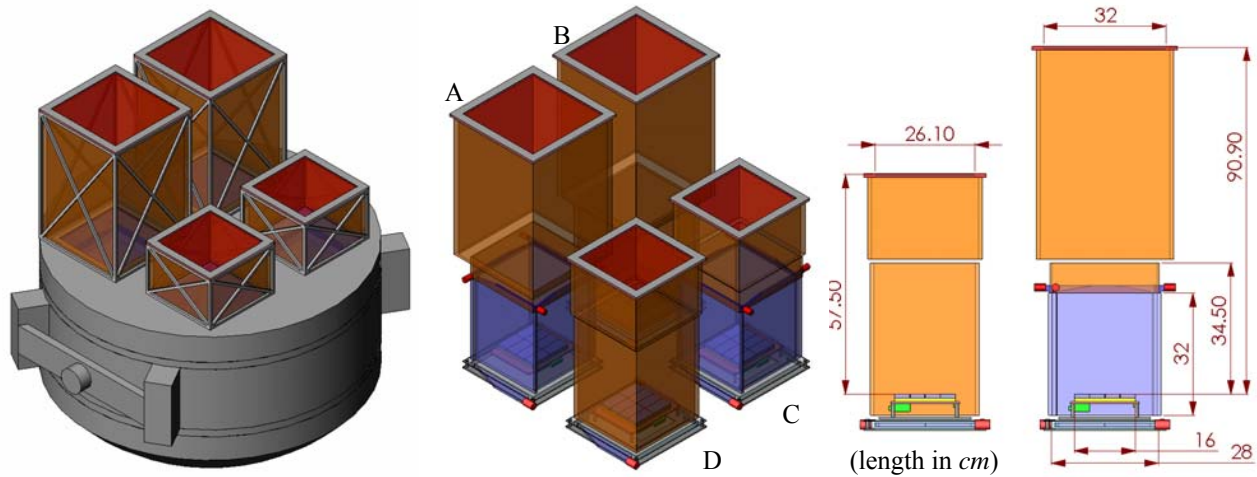

**Fig. 1.** *ProtoEXIST* Telescope configuration: overview of telescopes with the pressure vessel (left), an instrument-only view (middle), and an instrument-only side view (right). *ProtoEXIST1* will have four independent coded-aperture telescopes (see also Table 1). In order to reduce the size of the pressure vessel (1.2m diameter, ~50 cm tall cylinder, mounted on the elevation axis in the balloon gondola), the masks and the upper portion of (passive) side shields sit on top of the pressure vessel. The four telescopes have identical detector planes (16x16 cm$^2$ each) and CsI active rear shields (28x28 cm$^2$). The side shields have two configurations: A & C with the combination of CsI active (the lower portion) and Pb-Sn-Cu passive shield, and B & D simply with Pb-Sn-Cu passive shield. Each shield configuration has two different mask geometries, allowing two different FoVs (A & B vs. C & D). The gap in the side shields in the side view is due to the top cover of the pressure tank, but there is no direct path to the detector plane from stray background X-rays.

Hong et al. (2006) briefly introduced the overall concept of the telescope configuration for *ProtoEXIST1* [5]. Now we have solidified the design and are about to start building the mechanical structure. The main driver of the design concept is to identify the optimal telescope and shielding configuration for the HET in *EXIST* and to perform a rudimentary *scanning* operation using coded-aperture telescopes. Fig.1 shows the overview of the telescope configuration for *ProtoEXIST1*[*], which consists of four similar but disparate coded-aperture telescopes (marked by A, B, C and D). Table 1 summarizes the key parameters of the four telescopes in *ProtoEXIST1*.

All four telescopes will point at the same section of the sky and have the identical CZT detector planes (16x16 cm$^2$ each, 0.5 cm thick). This makes the comparison of the performance of the different shielding configurations relatively straightforward. The rear shield for each telescope is a CsI active shield (28x28 cm$^2$ each, 2 cm thick). This is necessary as the current background simulation study for *EXIST* indicates the rear active shield is essential due to large diffuse albedo X-rays from the atmosphere [6]. In addition, the extended spectral response (up to ~ 10 MeV) provided by the active shields is important for understanding the properties of GRBs, for which the active shields may be used as a calorimeter.

---

[*] *ProtoEXIST1* is for the detectors with 2.5 mm pixels, and *ProtoEXIST2* for 1.2mm pixels. See [5].

The active side shields also help to reduce the overall background and provide the additional spectral sensitivity for GRBs, but a key question to address in *ProtoEXIST* is whether the active side shield is necessary due to the complexity it may introduce to the HET in *EXIST*. If necessary, another question to address is the required height of such shielding. To investigate this issue, we employ the two types of side shield configurations: A & C with the combination of CsI active (the lower portion, 1 cm thick, extending ~30 cm above the detector plane) and Pb-Sn-Cu passive shields; vs. configuration B & D which is simply a Pb-Sn-Cu passive shield. Each of four side and one rear active shields will be made of a single CsI panel from ScintiTech[*]. The CsI shields will be read out by Photo Multiplier Tubes (PMTs) through light guides. We plan to use one PMT for each side shield for simplicity and two PMTs for the rear shield to increase the light collection efficiency.

In order to increase our understanding of the background for CZT detectors in a space environment, we allow two different types of field-of-views (FoVs, A & B vs. C & D), which are determined by the size of mask and passive side shields, which extend outside the pressure vessel (Fig. 1). This configuration will allow us to experimentally identify the diffuse X-ray background component from the internal background and vice versa. We expect two telescopes with the active side shields will have very similar internal background components, but the diffuse X-ray background will differ corresponding to the different FoVs. Similarly two telescopes with the same FoV will have the identical diffuse X-ray background, but the internal background will be different, depending on the presence of active side shields. Therefore, this two-by-two configuration will pin point each component of the background experimentally, thus helping to determine the optimal shield design for the HET in *EXIST*.

The masks will be made by laminating five layers of Tungsten plates (1 mm thick each) and the mask pattern will be generated by a laser etching technique. The mask pixel size is chosen to provide the similar positional resolution for all four telescopes. This is also for easy comparison of imaging performance among different configurations.

The four telescopes are spaced so that the FWHM of each FoV is not disrupted. In order to reduce the size of the pressure vessel (a 1.2m diameter, ~50 cm tall cylinder), the masks and the upper portion of (passive) side shields sit on top of the pressure vessel, over four X-ray entrance windows, which will be made of either thin carbon composite plate or thin Al foil supported by Kevlar strings. The overall instrument mass is estimated to be about 150 kg, and we expect the whole pressure vessel with the mask and their support structure to be about 300 kg. The pressure vessel also houses four electronic boxes – the flight computer module, the shield electronics module, the calibration source electronics box and power box.

**Table 1.** Key Parameters for *ProtoEXIST* Telescopes (variations in blue bold characters)

| Configuration | | A. Narrow-Active | B. Narrow-Passive | C. Wide-Active | D. Wide-Passive |
|---|---|---|---|---|---|
| Mask | | $32 \times 32 \times 0.5$ cm$^3$ W (0.5 cm pixel = $m$) | | **$26 \times 26 \times 0.5$ cm$^3$ W (0.25 cm pixel = $m$)** | |
| Mask – Detector ($f$) | | 91 cm | | **58 cm** | |
| Shield | Side | **CsI$^a$/Pb-Sn-Cu** | Pb-Sn-Cu | **CsI$^a$/Pb-Sn-Cu** | Pb-Sn-Cu |
| | Rear | $28 \times 28 \times 2.0$ cm$^3$ CsI | | | |
| Detector | | $16 \times 16 \times 0.5$ cm$^3$ CZT (0.25 cm pixel = $d$) | | | |
| FoV | Fully Coded | $10° \times 10°$ | | $10° \times 10°$ | |
| | FWHM$^b$ | $20° \times 20°$ | | **$25° \times 25°$** | |
| | FWZI$^c$ | $29° \times 29°$ | | **$37° \times 37°$** | |
| Angular Resolution$^d$ | | 21.1' | | 21.0' | |
| Instrument Mass | | 51kg | 42kg | 36kg | 23kg |

$^a$ $28 \times 40 \times 1.0$ cm$^3$, $^b$ Full Width Half Maximum. $^c$ Full Width Zero Intensity. $^d$ arctan[$(m^2+d^2)^{0.5}/f$]

---

[*] ScintiTech (http://www.scintitech.com/)

# 3. DETECTOR PLANE

(a) Detector Crystal Units (DCUs, 4 cm$^2$)

(b) Detector Crystal Arrays (DCAs, 32 cm$^2$)

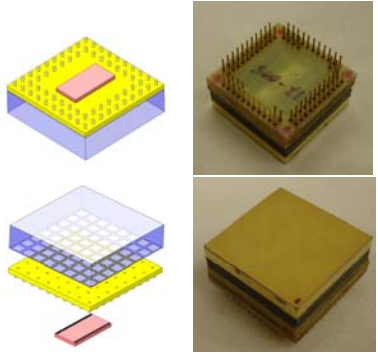
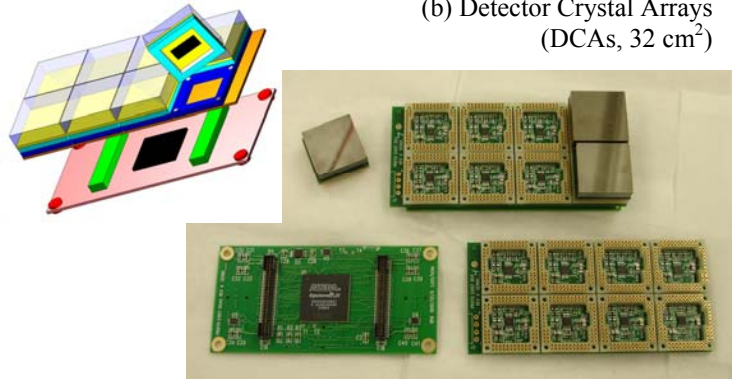

(c) Detector Module (DM, 256 cm$^2$) and FPGA Controller Board (FCB)

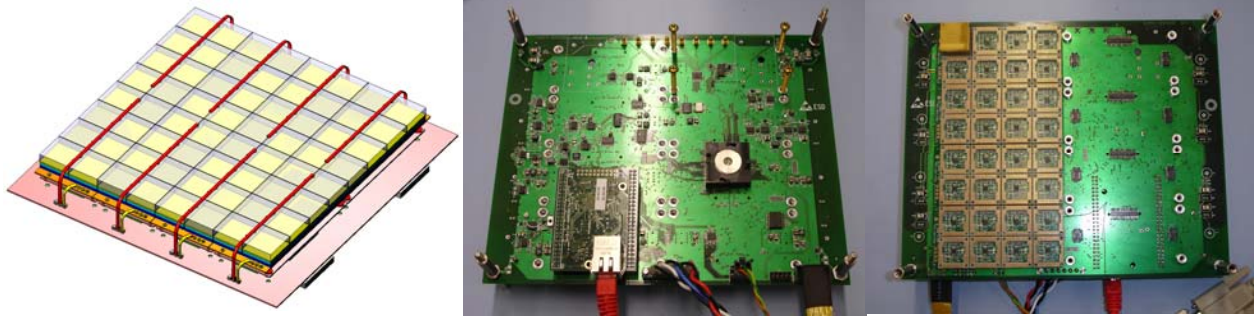

**Fig. 2.** Building a large area CZT detector plane. (a) Detector Crystal Unit (DCU, 4cm$^2$) concept and a Rev2 DCU, (b) Detector Crystal Array (DCA, 32cm$^2$) concept and Rev1 DCUs on a DCA, a DCA socket board, and a DCA FPGA board, and (c) Detector Module (DM, 256 cm$^2$) concept and the bottom side of a FPGA controller board (FCB) showing Netburner for Ethernet connection and FCB FPGA and the top side of the FCB mounted with four DCA boards (half-filled) and one Rev2 DCU.

Hong et al. (2006) describes our current detector packaging concept in detail [5]. Here we review the concept briefly and report recent developments. The actual hardware units of each detector plane, as shown in Fig. 2 have now been manufactured. The basic building block, the Detector Crystal Unit (DCU) is made of a 2 x 2 cm CZT crystal bonded on an interposer board (IPB), where a RadNET ASIC (for signal handling) is mounted [5, 7, 8]. We have assembled 85 Rev2 bare IPBs (pre-crystal bonding units: IPB + ASIC) from DDi[*] (and subcontractors), and recently completed a few Rev2 DCUs, which are currently undergoing a series of electronic pulser and radiation tests.

We have introduced a number of changes in Rev2 DCUs, compared to Rev1 DCUs in order to improve the spectral resolution [5]. First, for Rev1 DCUs we have used crystals from Orbotech Ltd[†] and eV products[‡]. For Rev2 DCUs, we employ crystals mainly from Redlen[§] Technology because of their low cost, uniform low leakage current (< 1nA per pixel), and ohmic IV characteristics under high voltage. Second, we have made a few changes in the design of the IPBs (e.g. board material changed from FR4 to Arlon 55NT, trace width and spacing from 5/5mils to 3/3 mils, etc) to reduce input capacitance noise and potential cross talks between pixels (see section 6 for the effects of the changes) [5]. Third, we have attached crystals on IPBs using a low-temperature solder bonding technique at Aguila Technologies[**]. The mechanical strength for the previous bonding method was provided by underfilling material, which contributes to input capacitance noise (~0.5 keV) [5]. In the Rev2 DCUs, we have achieved the same electric bond with no underfill material without sacrificing the mechanical strength of the bond. Fourth, in Rev2, the size of IPBs and crystals are trimmed or selected so that we can mount them closely next to each other (~ 400 um gap between crystals). We employ RadNET ASICs in both revisions [7, 8]; depending on the results of continuing tests, we will decide on the necessity of a revision for the ASIC for the next run.

---

[*] Dynamic Details, Inc (http://www.ddiglobal.com/)
[†] Orbotech Ltd (http://www.orbotech.com/)
[‡] eV products (http://www.evproducts.com/)
[§] Redlen Technology (http://www.redlen.com/)
[**] Aguila Technologies (http://www.aguilatech.com/)

We package a 2×4 array of DCUs onto a Detector Crystal Array (DCA). A DCA consists of two stacked circuit boards, the top board containing eight female sockets for mounting DCUs and the bottom board containing an Altera[*] Cyclone II FPGA to process signals from the ASICs of the eight DCUs. We have built ~ 10 pairs of DCA boards after a minor revision. The top-right panel in Fig. 2 shows Rev2 DCA modules along with Rev1 DCUs.

The Detector Module (DM) of each telescope in *ProtoEXIST* consists of 2×4 DCAs, which are mounted on an FPGA controller board (FCB). The bottom panel in Fig. 2 shows the DM concept and the bottom and top side of an FCB, where four DCA board pairs (and a Rev2 DCU) are mounted. The FCB also contains an FPGA of the Altera Cyclone II family to control and process signals from eight DCAs. The FCB hosts a NetBurner[†] card to communicate with the flight computer through Ethernet for data transfer and command upload. The FCB also allows eight TTL line inputs. Among them, five inputs are used for discriminator signals from five (or one) active shields, one input for the calibration source signals, and two inputs for time synchronization signals from GPS for 1 msec timing resolution. For calibration, we will use an $^{241}$Am source embedded in a plastic scintillator coupled to a PMT and mounted above each telescope.

The DCU/DCA mounting sockets on DCA boards/FCB are arranged so that the 20.40 mm pitch of the DCU sockets stays constant over the entire DM. This constant pitch is designed for optimal performance of coded aperture imaging, and it allows ~400 um gap between (Redlen) crystals. We will employ two EMCO[‡] HVPS units to apply the HV bias on 64 crystals (each for 32, not shown in Fig. 2). Each DCU will have a HV tab attached on the cathode surface (not shown in the figure). The tab wires will be bundled together and attached to a connector, which can be plugged in on the FCB. This biasing scheme is simplified from last year's concept [5] to avoid the complexity of the mechanical support system, while maintaining the capability of un-mounting/remounting of DCAs/DCUs if necessary. We allow a series of holes for swage nuts in FCBs for controlled mounting/un-mounting of DCAs from FCBs using a set of screws.

When an event occurs, the ASIC discriminator alerts the corresponding channel of the FPGAs in both the DCA and the FCB. Then, the DCA FPGA initiates the data readout sequence on the ASIC, while the FCB FPGA grabs the proper time and shield tag as well as the calibration source status. When the ASIC readout is finished, the FCB FPGA will combine the X-ray data packet from the DCA FPGA with the accessory data packet (time, shield, etc), and store them in the buffer that will be transferred via Ethernet (Netburner) to the flight computer for on-board processing and storage. The flight computer will combine the data from the 4 DMs (one for each telescope) for transmission to the ground.

## 4. PROGRESS IN DETECTOR PERFORMANCE

While we have successfully performed electronics pulser tests on multiple DCUs/DCAs on the FCBs using the FPGA code with limited features, the full operation (for pulser or radiation testing) is imminent. We are currently developing and optimizing the FCB FPGA codes for data taking and the data receiving clients for the flight computer.

In the meantime, for more efficient DCU testing and the verification of the functionality of DCAs, we have constructed two DCA test setups, one of which is shown in Fig. 3 along with the "poptart" system used for testing bare IPBs or DCUs [5]. The electronics pulser data indicate the noise performance of the bare IPBs on the DCA test system is consistent with the "poptart" system. We have been using both systems to test the DCUs and to finely tune the FPGA code for data taking. The "poptart" system is limited to operate one bare IPB or one DCU at a time due to the limitation of the FPGA code, but the DCA system can simultaneously operate up to eight bare IPBs or DCUs. The DCA test setup in Fig. 3 can apply HV bias on one DCU, but the 2$^{nd}$ DCA test setup can apply HV bias on all 8 DCUs when necessary (not shown in the figure). However, we plan to use a FCB test setup to calibrate all DCU units and to test the functionality of DCA boards in future. In the following, we compare the performance of the Rev2 vs. Rev1 DCUs based on the measurements performed in the "poptart" system and the DCA test system.

---

[*] Altera Corporation (http://www.altera.com/)
[†] Netburner Inc. (http://www.netburner.com/)
[‡] EMCO High Voltage Corporation (http://www.emcohighvoltage.com/)

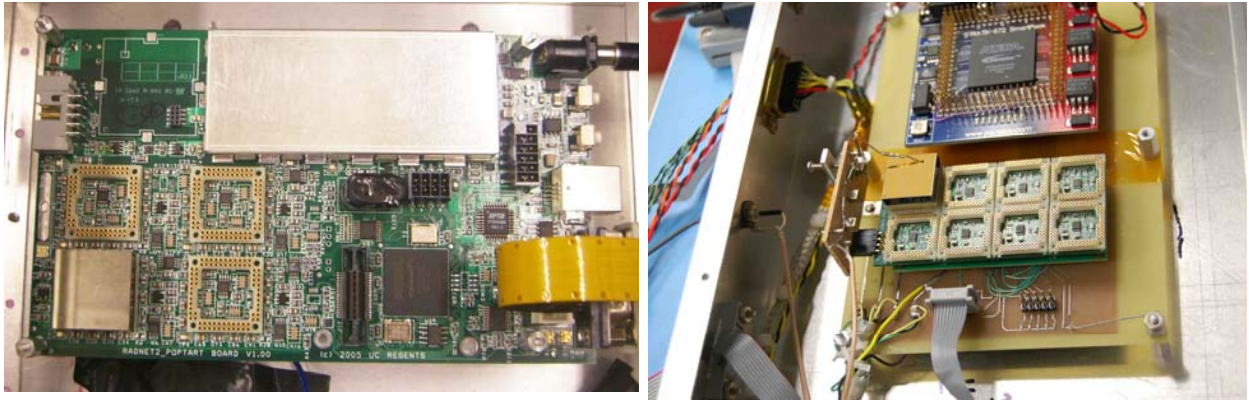

**Fig. 3** DCU test setup ("poptart" system, left) and DCA test setup (right). DCA test setups will be used to calibrate bare IPBs (pre-crystal bonding unit) and DCUs. The electronics pulser data indicate the noise performance of the bare IPBs on "poptart" systems is consistent with the same on DCA test systems.

### 4.1 Electronic Noise: Rev1 vs Rev2 bare IPBs

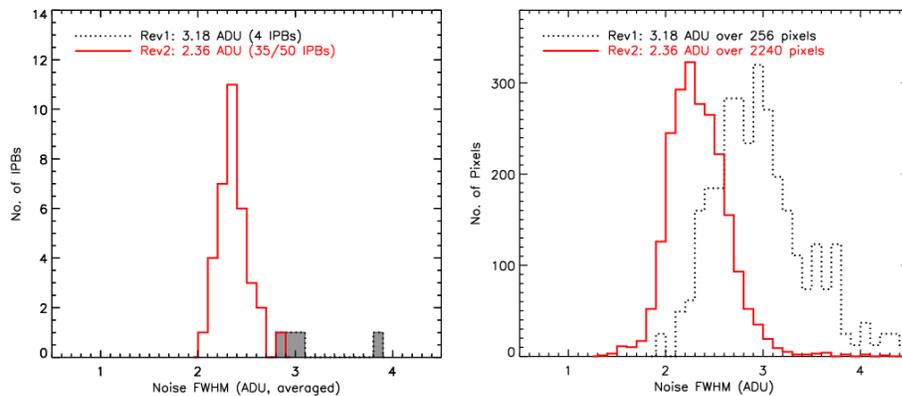

**Fig. 4.** IPB averaged electronics noise distribution of 35 Rev2 vs. 4 Rev1 bare IPBs (left), and the pixel-to-pixel noise distribution of the same (right), where the results for Rev1 are rescaled for easy comparison. On average, the noise of the 4 Rev1 IPBs is 3.18 ADU (FWHM), and the average of the 35 Rev2 IPBs is 2.36 ADU. Note 1 keV ~ 1.1 ADU.

Among the 85 Rev2 bare IPBs we have assembled, we have performed electronics pulser test on 50 IPBs so far. The 35 IPBs among these are internally graded A (70% yield), which means they have more than 60 working pixels with the electronics noise of 3.5 ADU[*] or less (FWHM). 30 of these have 64 working pixels and the remaining 5 have only one malfunctioning pixel.

Since we do not pre-screen ASICs (e.g. wafer level testing) before mounting, we estimate this 70% yield of bare IPBs directly reflects or closely follows the ASIC yield. In the case of Rev1 bare IPBs, we do not have a clear yield figure for either ASIC or IPB itself due to the low statistics and the nature of the initial development version. In the following we use the electronics pulser test data available on 4 Rev1 bare IPBs for comparison.

Fig. 4 shows the noise comparison (FWHM) of 4 Rev1 and 35 Rev2 bare IPBs. The left panel shows the IPB-averaged noise distribution and the right panel shows the pixel-to-pixel noise distribution. The results of Rev1 units in the right panel are rescaled for easy comparison. The improvement from Rev1 to Rev2 is ~0.75 keV (0.82 ADU) reduction in noise on average. We contribute this improvement to the reduced input capacitance by the change of the board material and trace thickness (section 4). Even the noisiest IPB of 35 Rev2 IPBs shows a slightly better performance than the 4 Rev1 IPBs.

---

[*] 1.1 ADU ~ 1 keV, varies with pixels.

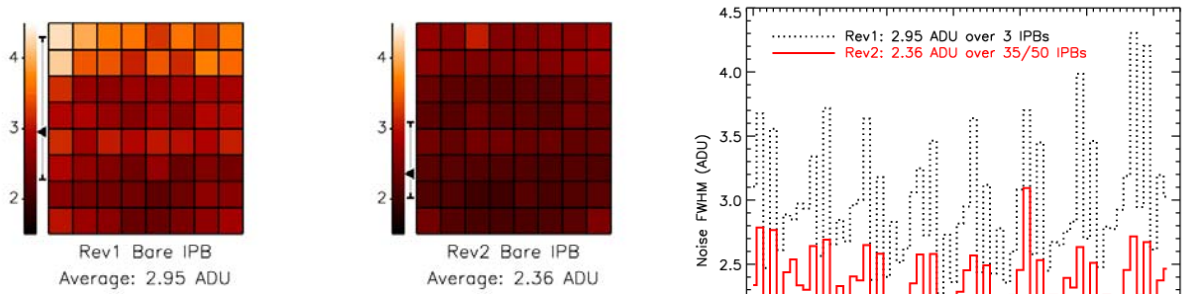

**Fig. 5.** Electronics noise distributions as a function of pixel, averaged over 3 relatively good Rev1 bare IPBs and 35 Grade-A Rev2 bare IPBs: 2-D pixel map geometry (left two panels) and pixel-to-pixel variation in 1-D plot (right). The Rev1 IPBs reveal the trace-length driven pattern more clearly (top two rows). The Rev2 IPBs also have this dependence, but the pixel-to-pixel variation is greatly suppressed.

Fig. 5 shows the same results as a function of the pixel (averaged over IPBs) to illustrate the pixel dependence of the noise. For easy comparison, we use 3 relatively good Rev1 bare IPBs. Both Rev1 and Rev2 bare IPBs show a clear pattern in the noise distribution, i.e. the top two rows of pixels in 2-D map have a larger noise than the rest of the pixels. This is due to the larger input capacitances caused by the longer trace lengths from those anode pixels to the ASIC input pads than the rest of the pixels [5]. In the case of the Rev2 bare IPBs, the variation among the pixels is dramatically suppressed, compare to the Rev1 units. Even if taking into account only 3 relatively good Rev1 IPBs, the improvement from Rev1 to Rev2 is clearly noticeable, and gives a ~ 0.5 keV reduction in noise.

### 4.2 Electronics Noise: Rev1 vs Rev2 DCUs

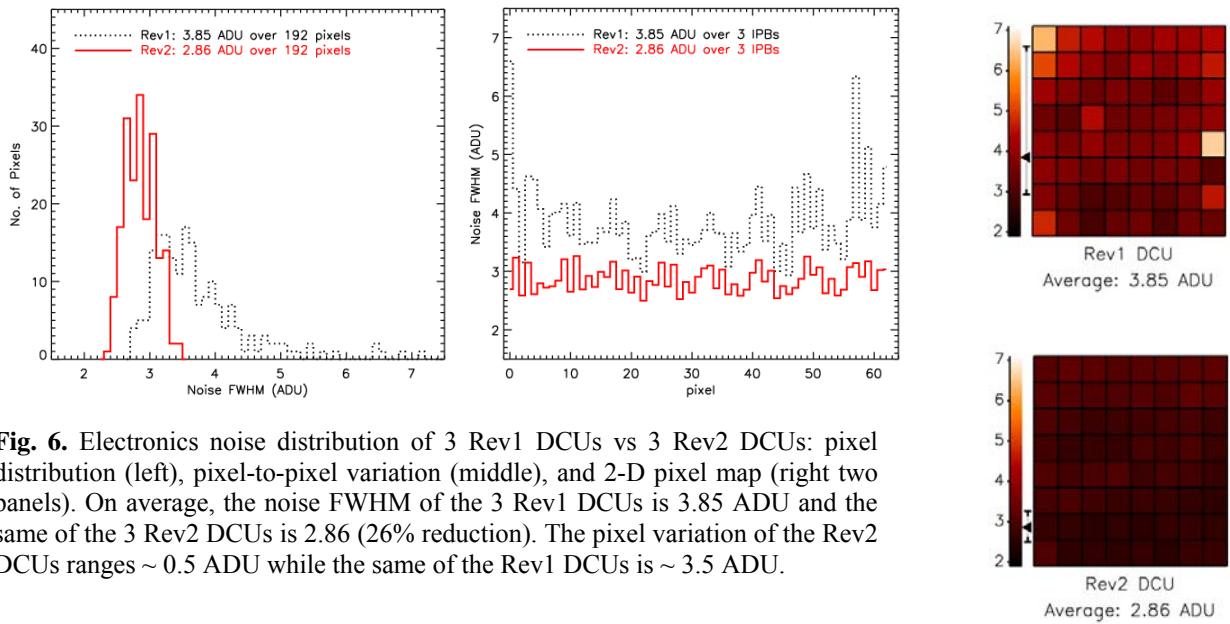

**Fig. 6.** Electronics noise distribution of 3 Rev1 DCUs vs 3 Rev2 DCUs: pixel distribution (left), pixel-to-pixel variation (middle), and 2-D pixel map (right two panels). On average, the noise FWHM of the 3 Rev1 DCUs is 3.85 ADU and the same of the 3 Rev2 DCUs is 2.86 (26% reduction). The pixel variation of the Rev2 DCUs ranges ~ 0.5 ADU while the same of the Rev1 DCUs is ~ 3.5 ADU.

Once a crystal is mounted on a bare IPB, the electronics noise of the unit increases due to the additional input capacitance introduced by bonding. In order to minimize this additional capacitance, in the Rev2 DCUs, we have removed the underfilling material for bonding (section 4). Fig. 6 shows a similar noise comparison between 3 Rev1 and

3 Rev2 DCUs. Note that these 3 Rev1 DCUs[*] were taken from a separate sample of bare IPBs different from the ones in Fig. 5, and the 3 Rev2 DCUs[†] were taken from the same sample as those in Fig.5.

In both cases the DCUs show an increase in noise as a result of bonding, but on average, the Rev1 DCUs show a larger increase (~0.82 keV) than the Rev2 DCUs (~0.45 keV). As a result, the performance gap between Rev1 and Rev2 further increases (~0.9 keV). The pixel-to-pixel variation is also greatly reduced (~0.5 ADU) in the Rev2 DCUs compared to the Rev1 DCUs (~ 3.5 ADU).

**4.3 Energy Resolution: Rev1 vs Rev2 DCUs**

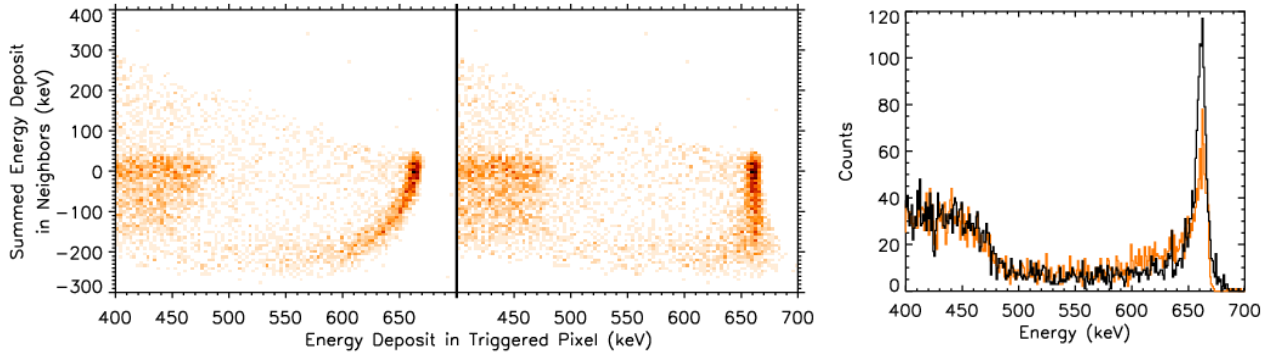

**Fig. 7.** The 662 keV line on a Rev1 DCU (1-050) from $^{137}$Cs. The left and middle panels show the summed signals of neighboring pixels vs. the signals from triggered pixels before/after the correction for depth-dependent incomplete charge collection (see [5]). The right panel compares the energy spectra for the 662 keV line before (orange) and after (black) the correction. The energy resolution at 662 keV is ~1.2% FWHM at 662 keV after correction.

Hong et al 2006 demonstrated the analysis technique using negative charges collected in the neighboring pixels, which allows both depth sensing and correction for depth-dependent incomplete charge collection [5, 9]. In the case of 1-050 (a Rev1 DCU), we have achieved 1.2% FWHM at 662 keV using this technique as shown in Fig. 7 (see [5] for the energy resolution of this detector at other energies). The left panel in Fig. 7 shows a scatter plot of events in the phase space of the summed signals from neighboring pixels vs. the signals from triggered pixels. The relative immobile holes induce predominantly negative charges in pixels neighboring triggered pixels, and the amount of the summed induced charges in the neighboring pixels depend on the depth of the events, so one can use this information to correct incomplete charge collection in the triggered pixels (middle panel in the Fig. 7). The resulting spectra before and after the correction are shown in the right panel in Fig. 7.

Based on the results in section 5.1 and 5.2, we expect a noticeable improvement in the energy resolution for the Rev2 DCUs compared to the Rev1 DCUs. In the case of the Rev2 units, we have started a series of radiation tests on the initial few units, and a detailed analysis is under way. Here we illustrate the spectral performance of the Rev2 DCUs, using the X-ray data from an $^{241}$Am source taken on 12-005 and we also introduce another correction for improving spectral resolution. Note that there are a few more corrections to apply for improvement beside the ones mentioned above. For example, charge splitting events due to charge spreading or Compton scattering [5] will be included.

The RadNET ASICs use a bank of 16 capacitors to sample 16 points of a pulse profile [5, 7, 8]. During normal operation, the ASIC is continuously sampling the signals from the pre-amplifier using these 16 capacitors in a circle, and when a trigger occurs, the ASIC holds the eight pre-trigger and eight post-trigger sampling values until they are read out (or the lock signal is released). The pulse height is given by the difference of the average of the last 5 post-trigger and the first 5 pre-trigger sampling values.

---

[*] 1-050, 1-003, 1-014
[†] 12-005, 12-020, 12-023

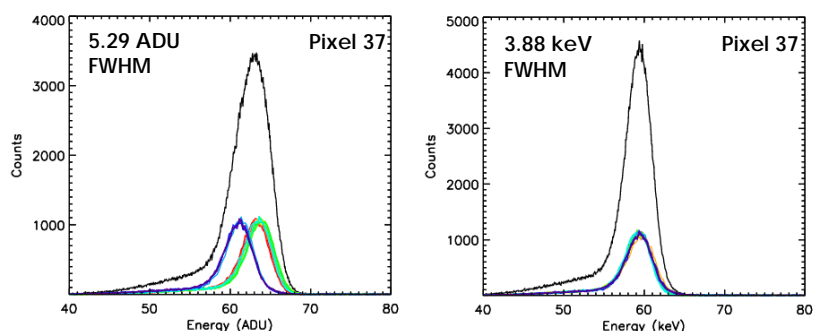

**Fig. 8.** Pulse height correction based on triggering capacitor IDs (CapIDs) (pixel 37 of 12-005). The (black) line is for the combined histogram, and the (colored) lines for histograms of each CapID. The left panel shows the raw histogram of the 60 keV line from an $^{241}$Am source, where two distinct groups of spectral distributions are present. The right panel shows the same after applying CapID dependent calibration. Pixel 37 shows ~23% improvement in energy resolution after the correction, but the whole DCU shows ~3% improvement on average (see the text).

An X-ray (or electronic test pulse) event can occur while any of 16 capacitors is sampling. Depending on which capacitor happens to sample at the moment of the trigger, the small difference in the 16 values of the capacitances causes variations in the final pulse height from the same energy deposit. This also causes null signal events produce non-zero pulse heights on average (either negative or positive). Note the RadNet ASICs can record null signal events even with a finite threshold by reading the signals from the pixels that are not involved in the event. This pulse height dependence on sampling sequence of capacitors also varies with pixels.

Fig. 8 shows the pulse height histogram of pixel 37 in 12-005 (a Rev2 DCU) recorded from 60 keV lines of the $^{241}$Am source. The plots show the combined (black) histogram of all events and the 16 (colored) histograms from each of the 16 triggering CapIDs. The summed histogram is binned in the ¼ scale compared to the histograms of the individual CapIDs for easy reading. The apparent wide spread of the pulse height histogram is partially caused by the small change in the pulse height depending on sampling sequence. As shown in this example, an anomaly of the 16 capacitors can cause two or three groups of pulse height histograms, depending on whether the particular capacitor is sampled pre-trigger, post-trigger or during the trigger. This can be corrected by generating separate conversion maps from ADU to energy (keV) for each of the 16 possible sequences. The result is shown in the right panel of Fig.8, showing ~23% improvement in the energy resolution.

In the case of 12-005, there are a few pixels such as pixel 37, which benefit significantly from this correction, but for the whole DCU, this correction only improves the resolution about 3% on average (from 4.09 to 3.98 keV). For most pixels, the pulse height variation due to sampling by a different sequence of capacitors (<~1keV) is much smaller than the overall energy resolution (~4 keV). For comparison, note that the incomplete charge correction by signals from neighboring pixels improves the resolution by ~ 6% at 60 keV in the case of 1-050 (a Rev1 DCU) [5].

Fig. 9 compares the energy resolution of 1-050 (Rev1 DCU) and 12-005 (Rev1 DCU). On average, there is about 0.7 keV reduction in energy resolution from 1-050 (4.7 keV) to 12-005 (4.0 keV). In this result, the correction for incomplete charge collection or charge splitting events is not applied for 12-005. Therefore, we expect the final resolution of 12-005 would be ~ 3.8 keV FWHM at 60 keV after all the corrections are applied. In the case of 1-050, the correction for different sampling sequence is not applied, but we estimate this correction will not have a noticeable impact due to the large intrinsic electronics noise in 1-050.

Table 2 summarizes the electronics noise and energy resolution budget on the Rev1/Rev2 bare IPBs/DCUs. The numbers in (red) bold are directly quoted from the actual measurements and the rest are simulation-based estimates or simple projections. We expect about 1.0% or less (FWHM) at 662 keV for Rev2 DCUs. Note that 1-050 is the best unit among the Rev1 DCUs we have tested, but we expect 12-005 to be a typical unit among Rev2 DCUs, judging from the uniformly high performance of the Rev2 bare IPBs and Redlen crystals we have measured.

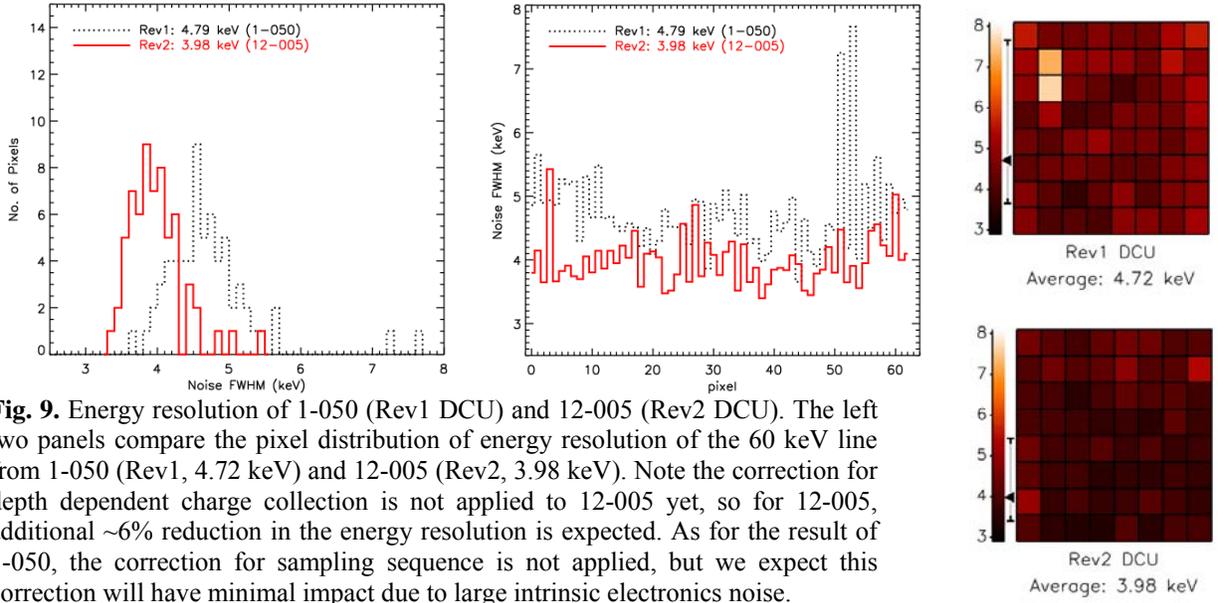

**Fig. 9.** Energy resolution of 1-050 (Rev1 DCU) and 12-005 (Rev2 DCU). The left two panels compare the pixel distribution of energy resolution of the 60 keV line from 1-050 (Rev1, 4.72 keV) and 12-005 (Rev2, 3.98 keV). Note the correction for depth dependent charge collection is not applied to 12-005 yet, so for 12-005, additional ~6% reduction in the energy resolution is expected. As for the result of 1-050, the correction for sampling sequence is not applied, but we expect this correction will have minimal impact due to large intrinsic electronics noise.

**Table 2** Electronics Noise and Energy Resolution Breakdown (FWHM in keV, actual measurements in bold red)

|  | Rev1 DCUs (3 IPBs, 1-050) | Rev2 DCUs (35 IPBs, 12-005) | DCUs for ProtoEXIST2[a] & EXIST |
|---|---|---|---|
| AISC intrinsic + wire bond | 1.5 – 2.0 | 1.5 – 2.0 | 1.3 – 1.7 (flip chip bond) |
| +Input traces | **2.9** | **2.0 – 2.8** | 1.3 – 1.7 (no IPB) |
| +Crystal bonding | **4.2** | **2.5 – 3.5** | 1.9 – 2.1 |
| +Leakage current, –600V | **4.5** | **3.6** | 2.0 – 2.4 |
| 59.5 keV ($^{241}$Am) | **4.7 (7.9%)** | **4.0 (6.7%)** | 2.5 (4.2%) |
| 122 keV ($^{57}$Co) | **5.5 (4.5%)** | 4.7 (3.8%) | 2.9 (2.3%) |
| 356 keV ($^{133}$Ba) | **7.8 (2.2%)** | 6.6 (1.9%) | 4.1 (1.2%) |
| 662 keV ($^{137}$Cs) | **7.9 (1.2%)** | 6.7 (1.0%) | 4.2 (0.6%) |

[a]1.2mm pixel detectors

## 5. SUMMARY AND FUTURE WORK

We have reviewed the telescope configuration for *ProtoEXIST*, and reported our progress of the development for CZT detectors. The telescope layout in *ProtoEXIST1* is designed to determine the optimal configuration for the HET in *EXIST*, and to demonstrate the performance of *scanning* coded-aperture imaging. At least a few units of each component of the detector plane hardware have been built and they are currently undergoing a series of electronics and radiation test. The Rev2 bare IPBs are being produced more reliably at high yield (~70%) than the Rev1 equivalent. A Rev2 DCU (12-005) shows the energy resolution of 4.0 keV or less (FWHM) at 60 keV, and is expected to show about 1.0% or less (FWHM) at 662 keV.

We are about to begin manufacturing the mechanical components for the pressure vessel and other telescope support structure. Another major task in addition to the hardware development is the FPGA and flight software development, which has just begun as of this writing. We also plan to develop the comprehensive calibration procedure and the automatic analysis software, which are essential for calibrating a large number of detectors (~300 DCUs for *ProtoEXIST1*).

## ACKNOWLEDGMENTS

This work is supported in part by NASA grants NAG5-5279, NAG5-5396 and NNG06WC12G. We dedicate this work to the late A. Capote at Aguila Technologies for his pioneering efforts and contributions to the project.

## REFERENCES

1. J. E. Grindlay et al., "EXIST: mission design concept and technology program," in *X-Ray and Gamma-Ray Telescopes and Instruments for Astronomy*; J. E. Truemper, H. D. Tananbaum; Eds., *Proc. SPIE* **4851**, pp. 331-344, 2003.
2. W. W. Craig, J. Hong and EXIST Team, "The Energetic X-ray Imaging Survey Telescope (EXIST): Instrument Design Concepts,"   AAS, **205**, 5001C, 2004
3. Grindlay, J. E. & Hong, J., "Optimizing wide-field coded aperture imaging: radial mask holes and scanning", *Optics for EUV, X-Ray, and Gamma-Ray Astronomy*, Edited by Oberto Citterio; Stephen L. O'Dell, Proceedings of the SPIE, Vol. **5168,** pp. 402-410, 2004.
4. A. Copete et al, in preparation, 2007
5. J. Hong et al., "CZT imaging detectors for *ProtoEXIST*" in *Hard X-Ray and Gamma-Ray Detector Physics and Penetrating Radiation Systems VIII*, Larry A. Franks; Arnold Burger; Ralph B. James; H. Bradford Barber; F. Patrick Doty; Hans Roehrig, Editors, *Proc. SPIE*, **6319**, 63190S, 2006
6. A. B. Garson et al "Background simulations for the energetic x-ray imaging survey telescope EXIST and the balloon-borne prototype experiment *ProtoEXIST*", This conference, *Proc. SPIE*, **6706**, 2007
7. W. R. Cook, J. A. Burnham and F. A. Harrison, "Low-noise custom VLSI for CdZnTe pixel detectors" in *EUV, X-Ray, and Gamma-Ray Instrumentation for Astronomy IX*; O. H. Siegmund, M. A. Gummin; Eds., *Proc. SPIE* **3445**, pp. 347-354, 1998
8. W. W. Craig, L. Fabris, J. Collins and S. Labov, "A Cellular-Phone Based Radiation Detector: Technical Accomplishments of the RadNet Program," UCRL-TR-215280.
9. J. D. Eskin, H. H. Barrett, and H. B. Barber, "Signals induced in semiconductor gamma-ray imaging detectors" J. Appl. Phys. **85**, pp 647-659, 1999